\documentclass[letterpaper]{article} 
\usepackage[]{aaai25}  
\usepackage{times}  
\usepackage{xcolor}
\usepackage{listings}
\usepackage{helvet}  
\usepackage{courier}  
\usepackage[hyphens]{url}  
\usepackage{graphicx} 
\usepackage{natbib}  
\usepackage{caption} 
\usepackage{algorithm}
\usepackage{algorithmic}

\usepackage{newfloat}

\DeclareCaptionStyle{ruled}{labelfont=normalfont,labelsep=colon,strut=off} 
\floatstyle{ruled}

\title{Keyword search is all you need: Achieving RAG-Level Performance without vector databases using agentic tool use}
\author{
    Shreyas Subramanian\equalcontrib,
    Adewale Akinfaderin\equalcontrib,
    Yanyan Zhang\equalcontrib,
    Ishan Singh\equalcontrib,
    Mani Khanuja,
    Sandeep Singh,
    Maira Ladeira Tanke
}
\affiliations{
    Amazon Web Services\\

    410 Terry Avenue North,\\
    Seattle, WA 98109\\
}

\usepackage{bibentry}

\begin{document}

\maketitle

\begin{abstract}
While Retrieval-Augmented Generation (RAG) has proven effective for generating accurate, context-based responses based on existing knowledge bases, it presents several challenges including retrieval quality dependencies, integration complexity and cost. Recent advances in agentic-RAG and tool-augmented LLM architectures have introduced alternative approaches to information retrieval and processing. 
We question how much additional value vector databases and semantic search bring to RAG over simple, agentic keyword search in documents for question-answering. 
In this study, we conducted a systematic comparison between RAG-based systems and tool-augmented LLM agents, specifically evaluating their retrieval mechanisms and response quality when the agent only has access to basic keyword search tools. Our empirical analysis demonstrates that tool-based keyword search implementations within an agentic framework can attain over $90\%$ of the performance metrics compared to traditional RAG systems without using a standing vector database. Our approach is simple to implement, cost effective, and is particularly useful in scenarios requiring frequent updates to knowledge bases. 
\end{abstract}

%

\section{Introduction}
The rapid advancement of Large Language Models (LLMs) has significantly transformed various industries and applications~\cite{hadi2023survey}, revolutionizing tasks such as text generation, summarization, question-answering systems, and chatbots~\cite{dam2024complete, kumar2023large}. 
These sophisticated models have demonstrated remarkable proficiency in understanding and generating human-like text, leading to widespread adoption in multiple sectors~\cite{hadi2023survey}. Recent developments in neural architectures and training methodologies~\cite{vaswani2017attention, brown2020language, liu2019roberta, raffel2020exploring, xu2022megatron} have enabled these models to process and generate increasingly complex and contextually relevant responses. 
However, as LLMs become integral to critical applications, the need for accurate information retrieval and response generation has intensified~\cite{bender2021dangers,van2024field}. This evolution has prompted a shift from standalone LLM responses to hybrid systems that integrate external knowledge sources, aiming to improve the factual accuracy and relevance of generated content~\cite{es2023ragas, zhao2024retrieval}. 

Retrieval Augmented Generation (RAG) systems have emerged as a prominent solution, combining LLMs with external databases to ground responses in factual information~\cite{lewis2020retrieval}. This approach mitigates issues like hallucinations (instances where models generate plausible but incorrect information) by providing access to up-to-date and domain-specific data. While RAG systems have proven effective, they face challenges in integrating retrieval mechanisms and maintaining knowledge bases. In response, tool-augmented LLM agents have been developed, using search engines and APIs to retrieve information dynamically, offering greater flexibility, particularly when knowledge requires frequent updates~\cite{qu2024tool}. Despite these advancements, systematic comparisons between RAG and tool-augmented approaches remain limited. Comprehensive benchmarking metrics assessing retrieval accuracy, response quality, latency, and maintenance overhead are essential for understanding their relative trade-offs and guiding the development of efficient, reliable LLM-based systems tailored to specific applications~\cite{gao2023retrieval, es2023ragas}.

This study aims to address this gap by conducting a comparison of traditional RAG systems vs. tool-augmented LLM agents. By systematically comparing their retrieval mechanisms and response quality, we seek to provide insights into their respective strengths and limitations. First, we cover a review of related work in the field of LLM-based retrieval systems and their applications. We then detail our methodology, which encompasses the implementation of RAG systems, tool-augmented LLM agentic frameworks, and our approach to evaluation using LLM-as-a-Judge. We then describe the datasets used in this study and our experimental setup. Finally we show how agentic systems without a standing vector database can perform competitively with traditional RAG approaches.

\section{Related Work}
\label{sec:relatedwork}

Retrieval-Augmented Generation (RAG) has emerged as a crucial approach for improving the accuracy and reliability of Large Language Models (LLMs) by combining retrieval mechanisms with generative capabilities~\cite{guu2020retrieval, yu2024defense, lewis2020retrieval}. By grounding responses in retrieved documents from external knowledge bases, RAG significantly reduces hallucinations and improves factual consistency in LLM outputs~\cite{hu2024refchecker, xu2024face4rag, zhao2024felm}. Studies have demonstrated that hybrid retrieval methods, incorporating both dense and sparse retrieval techniques, achieve superior performance in document retrieval quality~\cite{hambarde2023information, zhao2024dense}. Research by~\citet{rakin2024leveraging} shows that RAG systems utilizing dense passage retrieval can reduce model hallucinations significantly compared to base LLMs in fact-based question answering tasks. Additionally,~\citet{izacard2021leveraging} and~\citet{wang2023survey} found that the quality of retrieved passages directly correlates with the accuracy of generated responses, with their Fusion-in-Decoder approach demonstrating substantial improvements in response accuracy across various domains. These findings underscore RAG's significance in enhancing LLM performance through factual grounding and contextual relevance.

Tool-augmented Large Language Models (LLMs) represent an emerging alternative to traditional Retrieval-Augmented Generation (RAG) systems, offering dynamic access to information through integration with external tools such as search engines, APIs, and specialized databases~\cite{inaba2023multitool, prince2024opportunities, parisi2022talm}. Unlike static knowledge bases, these systems can interact with real-time data sources, allowing them to maintain accuracy in scenarios where information frequently changes~\cite{hong2024data}. The architecture of tool-augmented LLMs enables them to execute multi-turn interactions with various tools, significantly expanding their capability to provide contextually relevant and up-to-date responses. This approach proves particularly valuable in domains where information evolves rapidly, such as automated web navigation, automated game playing, database management, or scientific research, where traditional RAG systems might struggle to maintain current information without frequent updates to their knowledge bases~\cite{gur2023real, wang2023voyager, xu2023gentopia}.

The emergence of tool-augmented approaches has demonstrated significant effectiveness in handling complex queries that require current information, representing a notable advancement in LLM capabilities~\cite{xu2023gentopia, wu2024toolplanner}. These systems can dynamically select and utilize appropriate tools based on the query context, offering advantages in maintenance requirements compared to traditional retrieval-augmented implementations~\cite{mialon2023augmented}. The reduced need for maintaining extensive vector databases makes tool-augmented LLMs particularly attractive for resource-constrained applications, while their ability to execute real-time searches ensures accuracy in time-sensitive contexts. Furthermore, tool-augmented LLMs demonstrate enhanced adaptability to new scenarios and use cases, as they can leverage existing tools and APIs without requiring extensive retraining or knowledge base updates~\cite{mialon2023augmented}. This flexibility, combined with their ability to chain multiple tools together for complex reasoning tasks, positions tool-augmented LLMs as a promising direction for developing more versatile and maintainable AI systems~\cite{chen2024re}.

The comparative analysis of RAG and tool-augmented LLM approaches reveals significant gaps in systematic evaluation methodologies, particularly in standardized benchmarking datasets that reflect real-world document formats and retrieval scenarios. While both approaches demonstrate distinct advantages, the absence of comprehensive datasets that include varied document formats (PDFs, web pages, structured databases) hampers robust comparison~\cite{joshi2017triviaqa, kwiatkowski2019natural, pang2022quality, yang2018hotpotqa}. Current benchmarking efforts focus on retrieval accuracy, response quality, latency, and maintenance costs, but often fail to account for the practical challenges of handling diverse document types and formats~\cite{chen2024benchmarking, gao2023enabling}. This limitation is particularly notable in evaluating tool-augmented approaches that utilize document processing tools, where standardized testing frameworks are largely absent~\cite{yuan2024easytool}. Effective comparison requires multiple evaluation criteria, including retrieval precision, response accuracy, and user satisfaction, especially in user-facing applications where interaction quality and retrieval depth significantly impact system performance.

Operational considerations further highlight the distinct trade-offs between these approaches. RAG systems typically demand substantial resources for maintaining and updating knowledge bases, particularly challenging in rapidly evolving domains where frequent updates are necessary~\cite{guu2020retrieval, fan2024survey}. While tool-augmented LLM agents potentially reduce maintenance overhead by leveraging existing external tools, they introduce dependencies on external services and may face reliability issues when these services are unavailable~\cite{mialon2023augmented}. Vector database maintenance in RAG systems often incurs higher infrastructure costs, especially for organizations requiring frequent data updates~\cite{fan2024survey}. However, tool-augmented LLM approaches may encounter latency issues and operational risks due to external dependencies~\cite{qu2024tool}. These trade-offs become particularly crucial in applications where system adaptability and responsiveness are priorities, requiring careful consideration of resource management strategies and reliability requirements . The choice between RAG and tool-augmented approaches ultimately depends on specific use case requirements, available resources, and the balance between maintenance overhead and system reliability.

\section{Methodology}
\label{sec:method}

\begin{algorithm}[tb]
\caption{Agentic document based Question Answering}
\label{alg:algorithm}
\textbf{Input}: User query\\
\textbf{Parameter}: Folder with source files, Max iterations $t_{max}$\\
\textbf{Output}: Final answer
\begin{algorithmic}[1] 
\STATE Let agent iteration $t=0$.
\STATE Use \verb|pdfmetadata.sh| script to print metadata of all files in the folder
\WHILE{$t < t_{max}$}
\STATE Observe previous state
\STATE Write \verb|rga| or \verb|pdfgrep| or other linux command
\STATE execute command in linux shell
\IF {additional context found in observation}
\STATE Update answer
\ELSIF {Final answer found}
\STATE Stop searching 
\ELSE
\STATE Continue searching
\ENDIF
\ENDWHILE
\STATE \textbf{return} Final answer
\end{algorithmic}
\end{algorithm}

\begin{figure*}
    \centering
    \includegraphics[width=0.65\linewidth]{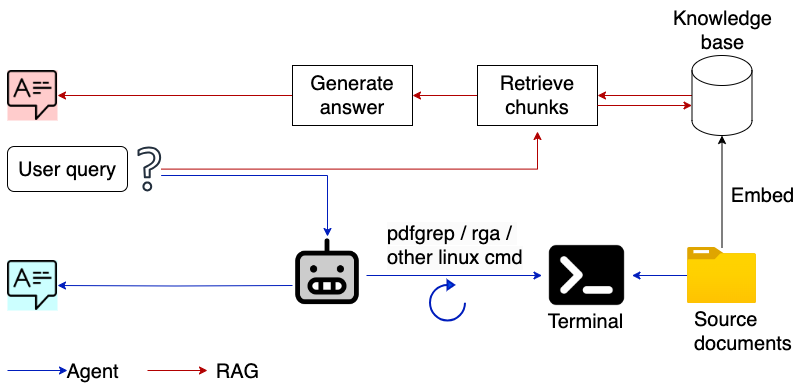}
    \caption{Comparison between RAG (\textcolor{red}{red}) and agent-based (\textcolor{blue}{blue}) pipelines for document QnA}
    \label{fig:enter-label}
\end{figure*}

In this study, we built a reference baseline vector database RAG that was used to compare the effectiveness of our agentic keyword-search approach. A high-level diagram of both approaches can be seen in Figure 1. Evaluations used LLM-as-a-Judge metrics computed using the RAGAS python library~\cite{es2023ragas} using datasets from various domains. Associated code is available in our github repository (see appendix).

\subsection{Datasets} To comprehensively compare traditional vector-based RAG approaches with our proposed agentic search methodology, we curated a diverse corpus of documents that vary in technical complexity, domain specificity, and linguistic structure. These datasets, sourced from llamahub~\cite{llamahub}, were selected as they represent standard RAG benchmarks and ensure reproducibility of our experiments. The datasets include the source pdf, along with questions and associated reference context and answers. The selected datasets represent some of the different challenges commonly encountered in real-world RAG applications: \begin{itemize} \item \textbf{PaulGrahamEssay:} The complete essays from Paul Graham's personal blog, selected for their complex argumentative structure and interdisciplinary nature~\cite{kamradt2023haystack}. \item
\textbf{Llama2:} The technical paper describing Meta's Llama 2 language model architecture and training methodology~\cite{touvron2023llama}. \item
\textbf{HistoryOfAlexnet:} Covers the development and impact of the AlexNet convolutional neural network~\cite{alom2018history}. \item
\textbf{BlockchainSolana:} Technical documentation and whitepapers related to the Solana blockchain platform~\cite{li2021bitcoin}. \item
\textbf{LLM:} A comprehensive survey paper on Large Language Models~\cite{guo2023evaluating}. \item
\textbf{FinanceBench:} A benchmark covering multiple publicly-traded company's public filings released between 2015 and 2023, including 10Ks, 10Qs, 8Ks, and Earnings Reports~\cite{islam2023financebench}. \end{itemize}

\subsection{Experiment 1: Baseline RAG Implementation} For our baseline comparison, we implemented a standard RAG pipeline using the fully-managed Amazon Bedrock Generative AI platform. An Amazon Bedrock Knowledge Base with Titan Text Embedding Model V2 with 1024 dimensional floating-point embeddings was used along with fixed 300 token chunking strategy with 20\% overlap. The source documents from all the datasets were ingested into an associated OpenSearch serverless index. The configuration used for retrieval include max number of chunk retrievals of 5. For response generation, the Anthropic Claude 3 Sonnet model with 200K context window and temperature set to $0.001$ was used with a RAG specific prompt detailed in the appendix. Results for the baseline RAG approach were generated for each dataset using the above retrieval and response pipeline with the corresponding dataset questions to be evaluated with ground truth answers and context.

\subsection{Experiment 2: Agentic Search Framework} 

Our proposed agentic approach leverages LLMs from the fully-managed Amazon Bedrock platform along with the open-source Langchain  framework. The LLM that we leveraged include the  Anthropic Claude 3 Sonnet hosted on Amazon Bedrock with a 200K context window.  We use use the standard  ReAct reasoning model~\cite{yao2022react} and set the temperature to $0.001$ for all experiments. The agent orchestration involves initial understanding of the query together and adopting a context-specific search strategy involving the following steps. The agent must begin with an initial metadata analysis of available documents in a folder. Then, the agent can dynamically decide to do broad keyword searches or targeted regex patterns across one or more documents. Using successive context expansion, more keyword searches, and error handling (for e.g. automatic retry) with modified search patterns the agent is able to perform a deep search via a linux shell; this is implemented using the LangChain experimental shell tool. For the agent to understand which documents may be relevant to the question, the agent is invoked with a custom prompt and directed to use several commands within the Linux shell based on the dynamic play between the initial query, search results, and alteration in approach as needed based on errors encountered or the outcomes of previous search iterations. The following are some of the commands the agent can execute: 

\begin{itemize} \item \emph{PDF Metadata Tool} that outputs metadata about the directory containing the files, and the individual files. \item \emph{RipGrep-All (rga)} for performing regex-based pattern matching, and multi-keyword search \item \emph{PDFGrep} providing PDF-specific search capabilities, page-range targeting and recursive directory search. \end{itemize} See Algorithm 1. for more details on the agent's search implementation. In addition, a detailed example agentic workflow can be found in the supplemental material.

Results are generated for each dataset by passing a subset of questions from the dataset to the keyword-search agent producing a candidate answer which can be compared with the ground-truth answer. The text segments selected by the agent from the source document were compared with the ground-truth contexts.

\subsection{Evaluation Methodology} 

To compare the baseline RAG approach with the keyword-search agent, the RAGAS evaluation framework~\citet{es2023ragas} was used. This provides a suite of traditional and LLM-as-a-judge based metrics. The primary metrics evaluated include: \begin{itemize} \item \textbf{Faithfulness} which measures factual consistency between the generated answer and the contexts used by the LLM to support its answer. \item \textbf{Context Recall} which measures the extent to which all the chunks relevant to answering a query are retrieved. \item \textbf{Answer Correctness} which measures the factual accuracy of the generated answer to the ground truth answer. \end{itemize} To assess variance in the agentic approach, metrics were calculated over multiple runs. For the FinanceBench dataset only answer correctness was evaluated and compared for the baseline RAG and agentic approaches due to the complexity of the tables and other complex structures present that poses a challenge for context-chunk based evaluation metrics.


\section{Results}
\label{sec:results}

\begin{table*}[t]
\centering
\begin{tabular}{l|l l c|l l c|l l c}
\hline
& \multicolumn{3}{c|}{\textbf{Faithfulness}} & \multicolumn{3}{c|}{\textbf{Context Recall}}  & \multicolumn{3}{c}{\textbf{Answer Correctness}} \\

\textbf{Dataset Name}  & Agent & RAG & Attain. (\%) & Agent & RAG & Attain. (\%) & Agent & RAG & Attain. (\%) \\
\hline
PaulGrahamEssay & 0.8662 & 0.9056 & 95.65 & 0.7527 & 0.8583 & 87.70 & 0.5808 & 0.7268 & 79.91 \\
Llama2Paper & 0.7252 & 0.8199 & 88.45 & 0.6148 & 0.8713 & 70.56 & 0.5823 & 0.6661 & 87.42 \\
HistoryOfAlexnet & 0.7280 & 0.7657 & 95.08 & 0.6968 & 0.8330 & 83.65 & 0.6406 & 0.7073 & 90.57 \\
BlockchainSolana & 0.8122 & 0.8627 & 94.15 & 0.7422 & 0.7450 & 99.62 & 0.5870 & 0.5872 & 99.97 \\
LLM Survey paper & 0.8061 & 0.8121 & 99.26 & 0.6355 & 0.6438 & 98.71 & 0.5123 & 0.5148 & 99.51 \\
\hline
\textbf{Average} & & & \textbf{94.52 \%} & & & \textbf{88.05 \%} & & & \textbf{91.48 \%} \\
\hline
\end{tabular}
\caption{Comparison of Agent vs RAG metrics across different datasets, including Attainment Scores (\%). Averages are shown in the final row.}
\label{tab:metrics_comparison_with_attainment_averages}
\end{table*}

The comparative analysis between the keyword search agent and baseline RAG approaches revealed interesting patterns across three key metrics: faithfulness, context recall, and answer correctness as detailed in Table 1. The agent's performance relative to RAG was evaluated using attainment scores, which represent the percentage achievement of the agent compared to the RAG baseline. In terms of faithfulness, the agent demonstrated strong performance, achieving an average attainment score of 94.52\% across all datasets. The agent performed particularly well on the LLM Survey paper dataset, reaching 99.26\% of RAG's performance, while showing slightly lower but still substantial attainment (88.45\%) on the Llama2Paper dataset. Context recall metrics showed more variability, with an average attainment score of 88.05\%. The agent achieved near-parity with RAG on the BlockchainSolana and LLM Survey paper datasets (99.62\% and 98.71\% attainment, respectively), though performance was notably lower for the Llama2Paper dataset (70.56\% attainment). For answer correctness, the agent maintained strong performance with an average attainment of 91.48\%. Particularly noteworthy were the results on the BlockchainSolana and LLM Survey paper datasets, where the agent achieved virtual parity with RAG (99.97\% and 99.51\% attainment, respectively). The lowest attainment in this category was observed in the PaulGrahamEssay dataset (79.91\%). For a coverage comparison of both approaches on all three metrics, see Figure 2.

\begin{table}[h]
\begin{tabular}{|l|c|}
\hline
System Configuration & Answer Correctness (\%) \\
\hline
Traditional RAG & 24.24 \\
Agent (3 run Average) & \textbf{32.71}\\
\hline
Agent (Run 4)$^*$ & 39.64 \\
\hline
\end{tabular}
\caption{Comparison of Agent vs RAG answer correctness (\%) across a subset of the FinanceBench dataset.
$^*$Run 4 removed Johnson \& Johnson reports from the dataset as the reference files were made unavailable publicly mid-experimentation.}
\end{table}

Overall, these results suggest that while the keyword search agent generally performed slightly below the RAG baseline, it maintained competitive performance levels, consistently achieving above 88\% average attainment across all three metrics without the use of any semantic search via a vector database. We specifically highlight the performance in the BlockchainSolana dataset where context recall and answer correctness are over 99\% of the baseline. This indicates that the agent-based approach could serve as a viable alternative to traditional RAG systems, particularly in scenarios where computational efficiency is desired or where vector-database usage is not optimal. Like all retrieval approaches, we observe performance variance across datasets and content types. For example, the PaulGrahamEssay dataset that is characterized by its interdisciplinary writing style, had lower attainment scores and underscores the importance of continued research into improving contextual comprehension beyond keyword matching.

 For the complex documents in the FinanceBench dataset, the results show consistent improvement over the traditional RAG baseline, with an average improvement of approximately 6 percentage points. The agent-based approach achieved a mean correctness score of $30.40\%$ ($\sigma = 1.31$), compared to $24.24\%$ for the traditional RAG system as detailed in Table 2. This improvement suggests that our agent's ability to actively search and interact with complex documents through commands provides more effective information retrieval compared to static chunk-based retrieval methods. The consistency across multiple runs also indicates the stability of our approach.

\subsection{Agentic keyword search vs. Claude computer use} 

Due to the recent popularity of \emph{Computer Use} capabilities introduced via beta features in Anthropic Claude Sonnet 3.5 v2 model~\cite{anthropic2024introducing} and more recently with OpenAI's operator~\cite{openai2025introducing}, we compared our method with this new capability that allows direct interaction with computer systems through shell commands and file operations, also offering a way to analyze documents and retrieve information without relying on vector databases~\cite{hu2024dawnguiagentpreliminary, lala2023paperqa}.
By using the Anthropic Claude Sonnet 3.5 v2 model, we created an agent that follows a structured workflow to answer questions based on access to the raw PDF datasets. 

We observe that the agent typically first opens the PDF in Firefox browser, uses keyboard shortcuts ($ctrl+F$) for precise term location, and captures screenshots of the relevant content before answering. We tested the Computer Use agent on seven representative questions from the FinanceBench dataset.\cite{islam2023financebench} and manually evaluated results . The agent consistently produced accurate and comprehensive answers. However, the approach faces certain operational challenges, common with our original agent, including occasional failures between API calls.  Side-by-side comparisons of our agent and computer use showed effectively similar results (see appendix), though we highlight that our agent approach is far simpler to set up and reproduce.

\begin{figure}[h!]
    \centering
    \includegraphics[width=0.85\linewidth]{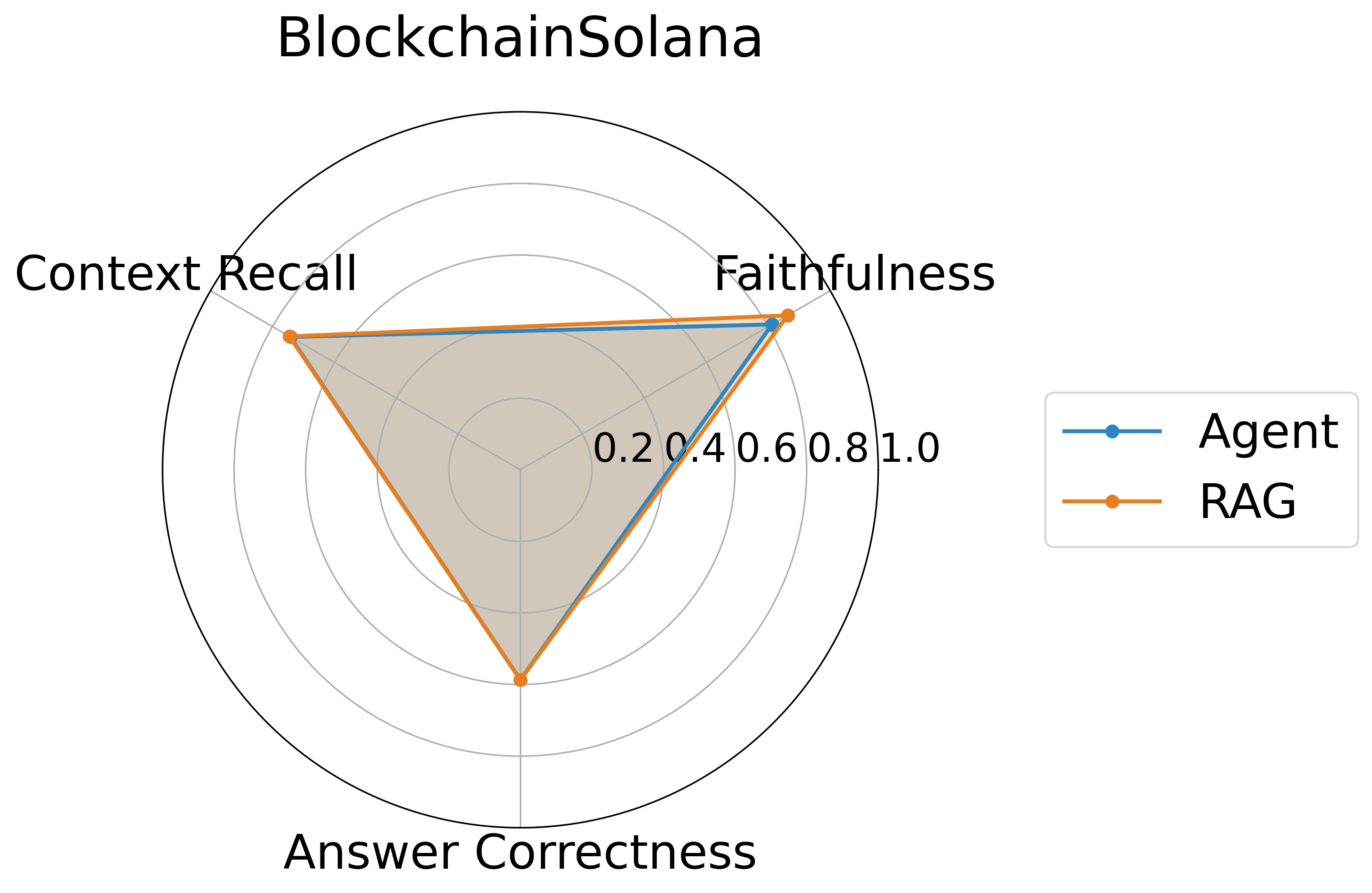}
    \includegraphics[width=0.85\linewidth]{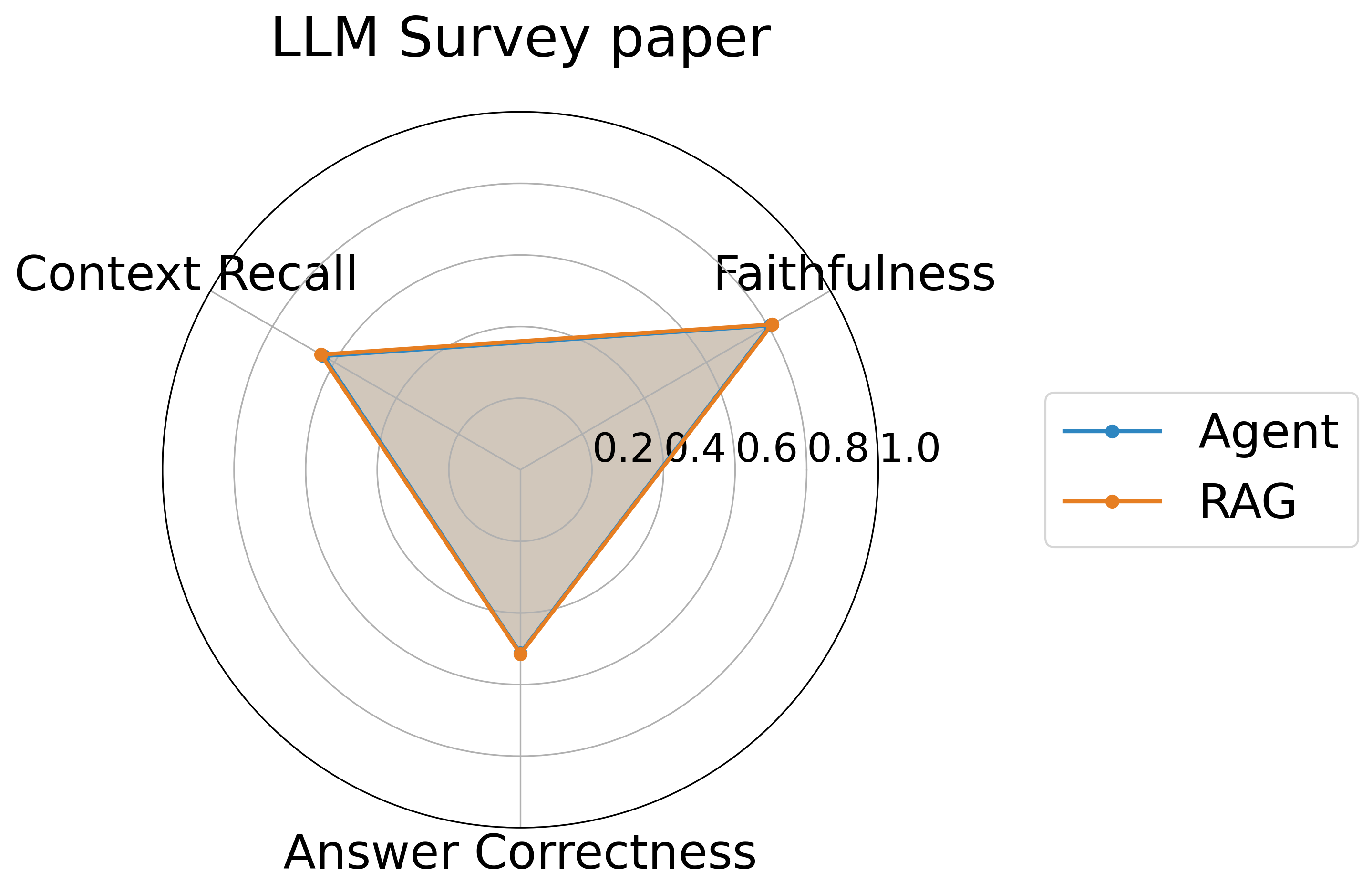}
    \caption{Coverage comparison of Tool-Augmented Agent vs RAG metrics across the BlockchainSolana and LLM Survey Paper datasets}
    \label{fig:enter-label}
\end{figure}

\section{Conclusion}
\label{sec:conclusion}
Our study demonstrated that agentic keyword search approaches can achieve comparable performance to traditional vector database RAG systems in document-based question answering tasks. Our experiments across diverse datasets showed that tool-augmented LLM agents using simple keyword search tools can attain over 90\% of the performance metrics of vector-based RAG implementations without the need for maintaining standing vector databases. By leveraging existing command-line tools and LLM reasoning capabilities, this method provides a robust alternative to traditional RAG systems, particularly in scenarios where information evolves rapidly or where resource constraints limit the feasibility of maintaining extensive vector databases.

However, several limitations were identified, including performance degradation with large documents, restricted multimedia handling, and context window constraints. The keyword search approach shows limitations in capturing contextual nuances, although partially mitigated through iterative refinement and the agent's inherent semantic reasoning capabilities. The current implementation also struggles with ambiguous queries and lacks long-term knowledge retention. Furthermore, privacy protections and ethical implications of automated retrieval systems require further investigation, including the implementation of appropriate guardrails such as data access controls and content filtering. Future research will focus on developing more automated, generalizable search strategies that can be useful across document types.

\section*{Appendix}

\subsection{Agent terminal tool instructions}

We use the standard zero shot agent template from langchain but provide a detailed description of the following tools it can use via the terminal:

\begin{lstlisting}
    terminal: Run shell commands on this Linux machine to search information in the "files/" folder. The commands to use are:


```pdfmetadata.sh```
--------------------
YOU MUST first print details of pdf files in the files/ folder (ALWAYS start with this without any changes)
# sh pdfmetadata.sh

This gives you file level metadata that is useful to narrow down the search. Then use rga or pdfgrep. Action input must start with rga or pdfgrep and contain the full command.:

```rga```
---------
A command line tool to search through files via keyword searches and regex patterns. All files relvent to this task are in the files/ folder.

- To find a search term in specific file (use regex pattern)
rga 'searchterm\w*' ./files/filename.pdf

- To search with multiple keyword matches across multiple files:
rga 'keyword1|keyword2|keyword3' ./files/

- Use -i for case insensitive search.


```pdfgrep```
-------------
Another commandline tool specifically for search with PDFs. Useful for special cases when:

- search in a folder with pdfs, across a specific page range for one or more keywords (-i is case insensitive, -n includes page numbers in output, -r is recursive search in files folder, and -P is perl compatible regex). With pdfgrep you must include '(' ')' brackets for the pattern
pdfgrep -inrP --page-range 1-4 '(keyword1|keyword2)' ./files/

- Search all .pdf files whose names begin with foo recursively in the current directory:
pdfgrep -r --include "foo*.pdf" pattern

Other tips:
***********
- If a complex query fails, try a series of simpler queries instead. 
- ALWAYS try to return larger context with -C 5 with both rga and pdfgrep to get 2 or more lines around the returned keyword match.
- remember that you MUST do 'sh pdfmetadata.sh' first to understand what files you are dealing with and then continue your search.
- your action should always just be "terminal" and action input is the full command you want to run in the terminal 
***********

\end{lstlisting}

\subsection{Example detailed agent run}

The table \ref{tab:my_label2} below shows a partial run of how our agent plans and executes commands to reach the final answer. The entire run can be seen in our supplementary material section of our associated github repository.\footnote{https://github.com/amazon-science/aws-research-science/tree/main/ShellAgent} For easier readability, commands run in the terminal corresponding to the original user question are shown in \textcolor{blue}{blue}. The agent first uses the `pdfmetadata' shell script to discover details about files existing in a local folder. Then, it uses `rga` to look for content related to Hyperledgerfabric. Then, it dives deeper into specific pages using the `pdfgrep' command to finally get the context required for the final answer.
\begin{table}[b!]
    \centering
    \begin{tabular}{p{2cm}p{6cm}}
        Step & Agent observation\\
        \hline
        \textbf{Question} & What are the three main components in Hyperledger Fabric and what role does each component play in the system?\\
        \textbf{Action input }& 
 \textcolor{blue}{sh pdfmetadata.sh} \\

\textbf{Observation} & -e File	Title	Author	Subject	Keywords	Creator	Producer	CreationDate	ModDate	Tagged	Pages	Encrypted	PageSize	FileSize	Optimized	PDFVersion
./files/BlockchainSolana.pdf		         Markus Richter	~~~ ...	\\
\textbf{Thought}   & The pdfmetadata output does not seem to contain information directly relevant to the main components of Hyperledger Fabric. I should search through the PDF files using the rga or pdfgrep tools to look for content related to "Hyperledger Fabric components", "Fabric architecture", or similar phrases.\\

\textbf{Action input} & \textcolor{blue}{rga -i 'hyperledger fabric components$\vert$fabric architecture' ./files/} \\

\textbf{Observation} & ./files/BlockchainSolana.pdf:Page 14: reflected in the increase in performance and strength on confidentiality. The Hyperledger Fabric architecture is shown in Fig. 9.
./files/BlockchainSolana.pdf:Page 15: Fig. 9. Hyperledger Fabric Architecture\\
 & The search results indicate that the PDF file "BlockchainSolana.pdf" contains information about the Hyperledger Fabric architecture and its main components. I should skim through the relevant pages to identify the three main components and their roles.
 \\
\textbf{ Action input} & \textcolor{blue}{pdfgrep -inrP --page-range 14-16 '(component$\vert$architecture)' ./files/BlockchainSolana.pdf -C 5} \\
 ... & (Results truncated here. Please refer to supplementary material.) \\
\textbf{Final Answer} & 
The three main components of Hyperledger Fabric are:

1. Membership - Provides identification services and allows nodes to access the system through membership services.

2. Blockchain - Provides consensus services for the distributed ledger. 

3. Chaincode - Refers to the programs or smart contracts that execute on the blockchain network.
     \end{tabular}
     \caption{Partial agent run showing how the original question is answered using keyword search tools.}
    \label{tab:my_label2}
\end{table}

\newpage
\subsection{Computer Use Agent Interactions}

Table \ref{tab:appendix_queries1} below shows example responses from our agent to various questions about American Express's 2022 financial statements, including the context used to derive each answer. These examples demonstrate our Computer Use agent's capabilities, which employs command-line tools and direct file operations to extract and verify information from PDF documents without relying on vector databases.

\begin{table}[h!]
    \centering
    \begin{tabular}{p{2cm}p{6cm}}
        \textbf{Question} & Which debt securities are registered to trade on a national securities exchange under American Express' name as of 2022?\\
        \textbf{Output} & As of 2022, American Express has only its Common Shares (par value \$0.20 per share) registered under Section 12(b) of the Securities Exchange Act to trade on the New York Stock Exchange under the symbol 'AXP'. No debt securities are registered to trade on any national securities exchange.\\
        \textbf{Context} & The company explicitly states under 'Securities registered pursuant to Section 12(b) of the Act' only its common shares. While American Express has significant debt securities outstanding, these are not exchange-listed. The company maintains debt ratings from major agencies (Moody's, S\&P, and Fitch) for its unsecured debt programs. Their debt securities are primarily issued through private/institutional channels rather than listed exchanges.\\
        \hline
        \textbf{Question} & What are the geographies that American Express primarily operates in as of 2022?\\
        \textbf{Output} & American Express operates globally with a strong presence in the United States and significant international operations. As of 2022, approximately 78\% of their total revenues net of interest expense were generated from U.S. operations, while 22\% came from international activities. Their largest international presence outside the United States includes the United Kingdom, Japan, the European Union, Australia, Canada, and Mexico.\\
        \textbf{Context} & The company operates in approximately 103 countries and territories through network partnerships. They have a regional structure divided into: United States, EMEA (Europe, Middle East and Africa), APAC (Asia Pacific, Australia and New Zealand), and LACC (Latin America, Canada and the Caribbean). International operations are managed through their International Card Services (ICS) segment.\\
        ... & (Results truncated here. Please refer to supplementary material.) 
    \end{tabular}
    \caption{Computer Use Agent Query Examples - Part 1}
    \label{tab:appendix_queries1}
\end{table}

\end{document}